# Skyrmion ratchet effect driven by a biharmonic force

Weijin Chen,[1,2,3] Linjie Liu,[2,3] Ye Ji[2,3] and Yue Zheng[2,3,*]

[1]*Sino-French Institute of Nuclear Engineering and Technology, Sun Yat-sen University, Zhuhai, 519082, China*

[2]*State Key Laboratory of Optoelectronic Materials and Technologies, Sun Yat-sen University, 510275, Guangzhou, China*

[3]*Micro&Nano Physics and Mechanics Research Laboratory, School of Physics, Sun Yat-sen University, 510275, Guangzhou, China*

Based on micromagnetic simulation and analysis of Thiele's equation, in this work we demonstrate that ratchet motion of skyrmion can be induced by a biharmonic in-plane magnetic field $h_x(t) = h_1\sin(m\omega t) + h_2\sin(n\omega t + \varphi)$, provided that integers $m$ and $n$ are coprime and that $m + n$ is odd. Remarkably, the speed and direction of the ratchet motion can be readily adjusted by the field amplitude, frequency and phase, with the maximum speed being over 5 m/s and the direction rotatable over 360°. The origin of the skyrmion ratchet motion is analyzed by tracing the excitation spectra of the dissipation parameter $\mathcal{D}$ and the skyrmion position **R**, and it shows that the dissipative force plays a key role in the appearance of ratchet motion. Such a ratchet motion of skyrmion is distinguished from those caused by single-frequency ac drives reported in the literature, and from that driven by pulsed magnetic fields as also predicted in this work. Our results show that skyrmion ratchet effect under biharmonic forces shares some common features of those found in many soliton systems, and the facile controllability of both the skyrmion speed and direction should be useful in practice.

[*] zhengy35@mail.sysu.edu.cn

I. INTRODUCTION

Since the theoretical prediction[1] and their first experimental evidence[2] made about a decade ago, magnetic skyrmions have become a constant focus of attention from both an academic and technological point of view. Magnetic skyrmions are a kind of whirl-like spin textures with typical sizes of 10-100 nm, and behave like stable particles under the protection of topology. Their presence, either as individual particles or in crystalline form (so-called Skyrmion lattice), has been identified in a series of bulk materials with chiral magnetism, as exemplified by B20 metal compounds (such as MnSi[3,4], MnGe[5,6], FeGe[7,8], $Fe_{1-x}Co_xSi$[9,10], $Mn_{1-x}Fe_xGe$[11], etc.), and multiferroic insulator $Cu_2OSeO_3$[12]. Magnetic skyrmions can be also stabilized in magnetic thin films contacted with heavy metal layers, such as Fe/Ir[13], Co/Pt[14], and CoFeB/Ta[15]. In most of the existing magnetic skyrmion systems, the emergence of the skyrmion is attributed to the Dzyaloshinskii-Moriya (DM) interaction[16,17], which arises from inversion symmetry breaking in the crystal lattice or at the interfaces. Due to the nontrivial topology, magnetic skyrmions have been shown to carry quantized emergent electromagnetic fields, which effectively act on conduction electrons and magnons, giving rising to intriguing physical behaviors in skyrmion systems such as topological Hall effects associated with the transport of skyrmions[18,19], electrons[20,21] and magnons[22,23]. In addition to the nontrivial topological behaviors, the charm of magnetic skyrmions also comes from their nanometric size[24], topological protection[25], ultralow electric currents required to drive their motion[26], and dynamics under microwave fields[27]. All these represent the importance of magnetic skyrmions in fundamental physics, and their high potential use of in future information memories and spintronic devices.

The understanding of the dynamics of a magnetic skyrmion in response to external sources is one of the important issues in the field of magnetic skyrmions, and is relevant for many applications. In particular, reliable control of skyrmion motion is a key for the application of racetrack-type skyrmionic devices. Previous works have shown that the translational motion of skyrmions can be driven by a variety of sources, which include time-unvarying sources like steady spin-polarized currents[28,29], electric field gradients[30], magnetic field gradients[31,32], and thermal gradients[31,32], as well as time-varying sources like single-frequency ac drives of currents, fields or field gradients[33-37].

For the cases of time-unvarying sources, skyrmion motion can be understood by the model that the skyrmion center is subjected to a steady driving force, arising from the spin transfer torque (STT), the spatial asymmetric potential due to field gradients, or the momentum transfer caused by the magnon flow, as is reflected by a steady and nonzero driving force **F** in Thiele's equation,

$$-\mathcal{M}\ddot{\mathbf{R}}+\mathbf{G}\times\dot{\mathbf{R}}-\alpha\mathcal{D}\dot{\mathbf{R}}+\mathbf{F}=0 \tag{1}$$

Here $\mathcal{M}$ is skyrmion mass, **R** is the collective coordinate of the skyrmion, $\mathbf{G} = \mathcal{G}\mathbf{e}_z = 4\pi Q\mathbf{e}_z$ is the gyromagnetic vector with $Q$ being the skyrmion charge, $\alpha$ is the Gilbert damping constant, and $\mathcal{D}$ is the dissipative force tensor. $\mathcal{D}$ is defined through the relation $\mathcal{D}_{ij} = \int(\partial_i\mathbf{m}\cdot\partial_j\mathbf{m})dxdy$, and it becomes $\mathcal{D}_{ij} = \delta_{ij}\mathcal{D}$ in the case of the highly symmetrical case of an isolated skyrmion.

Situations are more complicate for skyrmion motion under time-varying sources as excitation modes of skyrmion are involved. Several kinds of time-varying sources have been reported to induce a net unidirectional motion of skyrmion. The first way, as pointed out by Wang et al.[33], is via using an oscillating in-plane magnetic field together with a static in-plane magnetic field, which we would like to call a *biased* oscillating magnetic field. The key of this method is the breaking of the spatial symmetry of the skyrmion by the static in-plane field. The effect of such a biased oscillating magnetic field is the breaking of spatial symmetry of the force density and consequently, the time average of the driving force **F** in Thiele's Equation is nonzero. The second way reported by Moon et al.[34] is by using a *tilted* oscillating magnetic field (with in-plane and out-of-plane components). While the net motion of skyrmion can be intuitively understood by a mixing of the gyration and breathing modes, our analysis shows that the dissipation force term in Thiele's equation plays an important role in the net motion. The *tilted* oscillating magnetic field leads to an oscillation of the dissipation parameter $\mathcal{D}$ and the skyrmion velocity $\mathbf{v} = \dot{\mathbf{R}}$ with the same frequency, and consequently the time average of the dissipation force is nonzero. It is this net dissipation force that causes the net motion of skyrmion. The third way, as proposed by Reichhardt et al.[35,36], is to make use of oscillating drives like an ac current together with an asymmetry of the substrate. The net motion of skyrmion of this way is due to the asymmetric potential in space, and the Magnus force was shown to have a great impact on the skyrmion motion direction. The fourth way, which is realized by an oscillating magnetic field gradient of high symmetry, has been recently revealed by Psaroudaki and Loss[37]. For this case, the time-dependent dissipation terms caused by the coupling of external field with magnetic excitations is also the key for the appearance of unidirectional motion of skyrmion. It is noteworthy that such a motion exists even the system and the driving field are of high symmetry, in contrast to the previous cases.

From a fundamental point of view, the unidirectional motion of skyrmion under oscillating driving forces is relevant to a general kind of transport phenomena of soliton systems, named ratchet effect, where net motion of solitons is induced by zero-average forces. In the literature, the ratchet effect has been explored in many different soliton systems by physicists and mathematicians. In particular, a large number of works have shown that a ratchet effect would appear if the system is driven by a biharmonic force $f(t) = f_1\sin(m\omega t) + f_2\sin(n\omega t + \varphi)$[38,39], and the speed and direction can be readily adjusted by the frequency $\omega$ and phase $\varphi$, provided that $m$ and $n$ are two coprime integers such that $m + n$ is odd. It is natural to ask if magnetic skyrmion exhibit a ratchet effect under a biharmonic driving force. The existence of such a ratchet effect can not only provide us an alternative strategy to control skyrmion transport, but also help us to gain a deeper insight into the skyrmion dynamics. Nevertheless, explorations on skyrmion dynamics under a biharmonic force remain exclusive.

In this paper, we study the skyrmion dynamics under biharmonic magnetic fields. Based on micromagnetic simulation and analysis of Thiele's equation, we show that a ratchet motion of skyrmion can indeed be induced by a biharmonic in-plane magnetic field $h_x(t) = h_1\sin(m\omega t) + h_2\sin(n\omega t + \varphi)$, provided that $m$ and $n$ are two coprime integers such that $m + n$ is odd. The direction of the motion can be continuously rotated by 360° by adjusting the phase $\varphi$. The ratchet motion speed can be tuned by both frequency and the field amplitude and is most significant near the resonance frequency of the gyration mode. The analysis on Thiele's equation shows that the appearance of a net dissipative force, due to an overlapping of the excitation modes of the dissipation parameter $\mathcal{D}$ and the skyrmion poison **R**, is believed to be the key to the ratchet motion. The difference between such a ratchet effect and those driven by single-frequency oscillating driving forces as well as that caused by pulsed magnetic fields as predicted at the end of this work is discussed.

## II. MODEL AND METHOD

The motion of an isolated skyrmion in a chiral magnet with bulk DM interaction is numerical studied. Basing on a Heisenberg model on a two dimensional square lattice, we write the following effective Hamiltonian of a chiral magnet as[40]

$$\mathcal{H}(\mathbf{m}_i) = -J\sum_{<i,j>} \mathbf{m}_i \cdot \mathbf{m}_j - D\left(\sum_i \mathbf{m}_i \times \mathbf{m}_{i+e_x} \cdot \hat{e}_x + \sum_i \mathbf{m}_i \times \mathbf{m}_{i+e_y} \cdot \hat{e}_y\right) - \sum_i \mathbf{H}(t) \cdot \mathbf{m}_i \quad (2)$$

where $\mathbf{m}_i$ is the magnetization vector at site *I*, *J* is the Heisenberg exchange coefficient, *D* is the DM interaction coefficient, and $\mathbf{H}(t) = \mathbf{H}_0 + \mathbf{h}(t)$ is the external magnetic field which is the sum of a constant field normal to the plane $\mathbf{H}_0=(0, 0, H_z)$ and a time-varying in-plane field $\mathbf{h}(t)$.

The dynamics of skyrmion is captured by solving the stochastic Landau-Lifshitz-Gilbert (LLG) equation,

$$\frac{d\mathbf{m}_i}{dt} = -\gamma\left[\mathbf{m}_i \times \left(\mathbf{H}_i^{\text{eff}} + \mathbf{L}_i^{\text{fl}}(t)\right)\right] + \alpha\left(\mathbf{m}_i \times \frac{d\mathbf{m}_i}{dt}\right) \quad (3)$$

or in the equivalent form,

$$\frac{d\mathbf{m}_i}{dt} = -\frac{\gamma}{\alpha^2+1}\left[\mathbf{m}_i \times \left(\mathbf{H}_i^{\text{eff}} + \mathbf{L}_i^{\text{fl}}(t)\right) + \alpha \mathbf{m}_i \times \left[\mathbf{m}_i \times \left(\mathbf{H}_i^{\text{eff}} + \mathbf{L}_i^{\text{fl}}(t)\right)\right]\right] \quad (4)$$

where $\gamma = g\mu_B/\hbar$ is the gyromagnetic ratio, $\alpha$ is the Gilbert damping coefficient, $\mathbf{H}_i^{\text{eff}}$ is the effective magnetic field given by $\mathbf{H}_i^{\text{eff}} = -\partial \mathcal{H}/\partial \mathbf{m}_i$, and $\mathbf{L}_i^{\text{fl}}(t)$ is the stochastic field caused by the effects of a thermally fluctuating environment interacting with $\mathbf{m}_i$. $\mathbf{L}_i^{\text{fl}}(t)$ satisfies $\langle \mathbf{L}_i^{\text{fl}}(t) \rangle = 0$ and $\langle L_{i\beta}^{\text{fl}}(t) L_{i\lambda}^{\text{fl}}(s) \rangle = \alpha k_B T \gamma^{-1} m^{-1} \delta_{ij} \delta_{\beta\lambda} \delta(t-s)$, where $\beta$ and $\lambda$ are Cartesian indices, $T$ is temperature, $k_B$ is the Boltzmann constant, and $m = |\mathbf{m}_i| = |g\mu_B|/a^3$ is the norm of the magnetization vector.

Numerical simulations based on the stochastic LLG equation are performed via an explicit Euler iteration scheme. The size of sample systems is fixed to be $128 \times 128$ sites under the periodic boundary condition. The Heisenberg exchange $J$ is taken to be $J/k_B$ = 50 K[32], and the strength of the DM interaction coefficient is $D = 0.15J$. The spin turn angle $\theta$ in the helical structure is ~6° as determined by $\theta = \arctan[D/(\sqrt{2}J)]$[40]. This results in the skyrmion diameter of ~30 nm if we consider a typical lattice parameter of $a = 5$ Å, as shown in Fig. 1a. The Gilbert damping coefficient $\alpha$ is taken to be 0.1. The external magnetic field normal to the plane is fixed to be $H_z = 0.01$, in unit of $J/(g\mu_B)$, which is ~0.11 T for $g$ equal to 6.74. The time step is taken to be 0.01, in unit of $\hbar/J$, which is ~1.5 fs. The in-plane magnetic field takes a biharmonic form along $x$ direction $h_x(t) = h_1\sin(m\omega t) + h_2\sin(n\omega t + \varphi)$. To first obtain the steady skyrmion, the magnetic structure is initially set with a downward magnetization in the center region and with an upward magnetization elsewhere, and is relaxed over a sufficiently long time (>3 ns). In the following, we focus on the results obtained at 0 K. A finite temperature would not change the main conclusions of this work.

To characterize the skyrmion, we calculate the topological charge density,

$$q = \frac{1}{4\pi} \mathbf{m} \cdot \left( \partial_x \mathbf{m} \times \partial_y \mathbf{m} \right) \tag{5}$$

as defined in the continuous form. The distribution of the topological charge density of a skyrmion at static state is shown in Fig. 1b. The total topological charge is then given by

$$Q = \int q \, dx \, dy \tag{6}$$

And the position of skyrmion $\mathbf{R} = (x_c, y_c)$ can be determined by

$$x_c = \frac{\int xq \, dx \, dy}{Q}, \quad y_c = \frac{\int yq \, dx \, dy}{Q} \tag{7}$$

## III. RESULTS AND DISCUSSIONS

**A. Excitation-spectrum of a skyrmion under in-plane magnetic field**

We would like to first take a look at the excitation-spectrum of a skyrmion caused by in-plane magnetic fields. Such a spectrum is obtained by tracing the spin dynamics of the skyrmion after applying a $\delta$-function pulse of in-plane magnetic field $h_x(t) = h_0 \delta(t)$ at $t = 0$. The time evolution curves of the skyrmion position coordinates $x_c$ and $y_c$ are shown in Fig. 1c, and the trajectory of the skyrmion after excitation is also depicted in the insert. One can see that the pulsed in-plane field drives the skyrmion into a damping counterclockwise (CCW) gyration around the equilibrium position ($64a$, $64a$). Based on the Fourier transformations of the position coordinates $x_c$ and $y_c$ of the skyrmion, we calculate the excitation-spectrum of the skyrmion by the power spectra of $x_c^*(\omega)x_c(\omega)$ and $y_c^*(\omega)y_c(\omega)$ as shown in Fig. 1d. A resonant frequency $f_r$ of about 16.4 GHz is clearly seen, corresponding to the CCW gyration mode of the skyrmion. Note that the gyration trajectory of skyrmion under a harmonic in-plane magnetic field is generally ellipse rather than a circle, reflecting the fact that the trajectory is actually a supposition of the CCW and the CW gyration modes[41]. A net CW gyration of the skyrmion is not seen as the resonant frequency of the CW gyration mode is about zero and its amplitude is always smaller than that of the CCW gyration mode for a non-bounded free skyrmion. The resonant frequency provides us an estimation of the skyrmion mass $\mathcal{M} \sim -\mathcal{G}/\omega_r$ ~ 0.122 ns, where $\omega_r = 2\pi f_r$ is the resonant angular frequency.

## B. Skyrmion dynamics under biharmonic magnetic fields

We then study the skyrmion dynamics under biharmonic in-plane magnetic fields along x-axis in form of $h_x(t) = h_1\sin(m\omega t) + h_2\sin(n\omega t + \varphi)$. For simplicity, in the work we set $h_1 = h_2$. Fig. 2a and b respectively depict the snapshots of skyrmion configuration under a period of two specific biharmonic fields, $h_x(t) = 0.003[\sin(\omega t) + \sin(3\omega t)]$ and $h_x(t) = 0.003[\sin(\omega t) + \sin(2\omega t)]$, with the angular frequency of the fields $\omega = 105$ rad/s, and the corresponding frequency $f = 16.7$ GHz, and period $T = 60$ ps. That is to say, the frequencies of the two harmonic magnetic field components have the relation $m + n =$ even for the first case, and $m + n =$ odd for the latter case. From the field profiles, one can also note that the first field $(m, n) = (1, 3)$ is time symmetric with $h_x(t+T/2) = -h_x(t)$, whereas the second field $(m, n) = (1, 2)$ does not have such a symmetry. The plot settings of the skyrmion configurations are the same as that of Fig. 1a, with an arrow plot of the magnetization vectors projected onto the xy plane at sites $(i_x, i_y)$ satisfying $\mod(i_x, 2) = \mod(i_x, 2) = 0$ and a color map of the z-axis component magnetization $m_z$. For each case, the snapshots are taken at five time points during a time period from time point $20T$ to $21T$ as labeled in the field profiles. Such a time interval is chosen to guarantee that the skyrmion is already in a steady excitation. The skyrmion dynamics of the two cases are quite similar at the first sight on the snapshots of skyrmion configuration. For both cases, the skyrmion deforms and gyrates in CCW direction. However, by tracing the skyrmion position, we found that in the first case, the skyrmion exhibits a close trajectory after a period of field application, whereas for the latter case, the trajectory is not closed after a period of field application, with a net displacement of the skyrmion position (1→5). Since all the conditions return to be the same except for the skyrmion position after a period, one expects that the drifting of the skyrmion in the latter case along the direction (1→5) will be accumulated if one traces the skyrmion motion over more time periods. That is to say, a ratchet motion occurs.

To clearly see the difference between the skyrmion dynamics driven by these two biharmonic in-plane magnetic fields, in Fig. 3 we further show the long-time skyrmion dynamics under the two fields up to 1.5 ns. The field profiles, the time evolution curves of the skyrmion position coordinates $x_c$ and $y_c$, and the trajectories of the skyrmion for the two cases are shown in Fig. 3a to c, and Fig. 3d to f, respectively. It is clearly seen that the skyrmion in the first case indeed performs a bounded periodic motion around the equilibrium position at rest, whereas the skyrmion in the latter case shows a ratchet motion with a helical-like motion trajectory along a specific angle direction. The ratchet motion speed of the skyrmion, which can be calculated as $v_c = \sqrt{v_x^{c2} + v_y^{c2}}$ [with $v_x^c = \lim_{t\to\infty}\left(\frac{1}{t}\int_0^t x_c dt - x_c^0\right)/t$, $v_y^c = \lim_{t\to\infty}\left(\frac{1}{t}\int_0^t y_c dt - y_c^0\right)/t$, and ($x_c^0$, $y_c^0$) being the initial position of the skyrmion], is found to be about $9.5\times10^9$ $a$/s, and the ratchet motion direction, which is defined as $\theta = \arctan\left(v_y^c / v_x^c\right)$, is found to be about 313°. If we take

$a$ = 0.5 nm, the ratchet motion speed is about 5 m/s. This value is smaller to that driven by current[28,29], but is comparable to those driven either by using an biased oscillating magnetic field[33] or by using a tilted oscillating magnetic field[34] as predicted in previous works.

To confirm the occurring condition of skyrmion ratchet motion under biharmonic in-plane magnetic fields, we further explore the skyrmion dynamics under biharmonic in-plane magnetic fields $h_x(t) = 0.003[\sin(m\omega t) + \sin(n\omega t)]$ with other values of coprime integers $(m, n)$. We still take $\omega$ = 105 rad/s. The long-time skyrmion motion trajectories of cases $(m, n)$ = (2, 3), (1, 4), (1, 5), and (2, 5) up to 1.5 ns are shown in Fig. 4a to d, respectively. From these trajectories, we find that ratchet motion of skyrmion occurs for cases $(m, n)$ = (2, 3), (1, 4), and (2, 5), whereas the skyrmion in case $(m, n)$ = (1, 5) exhibits a bounded motion. These results, together with the previous two cases with $(m, n)$ = (1, 2), and (1, 3), clearly show that a ratchet motion is induced when the frequencies of the two harmonic field components satisfy the relation $m + n$ = odd, otherwise a bounded motion would appear when $m + n$ = even. That is to say, despite of the different physics and equations that describe the systems, dynamics of magnetic skyrmion under a biharmonic force does share a generic feature found in many other soliton systems[36,37]: a ratchet motion can appear if some temporal symmetries are broken by time-dependent forces.

## C. Tunability of the skyrmion ratchet motion

In this section, we would like to show the facile tunability of the skyrmion ratchet motion under biharmonic in-plane magnetic fields. For biharmonic in-plane magnetic fields in form of $h_x(t) = h_x[\sin(m\omega t) + \sin(n\omega t+\varphi)]$, besides the relation between $m$ and $n$ which determines the occurring of ratchet motion, one can rely on the field amplitude $h_x$, frequency $\omega$ and phase $\varphi$ to tune the ratchet motion. As an example, we illustrate the case $(m, n) = (2, 1)$. The results of cases with other values of $(m, n)$ should be similar. The dependence of the ratchet motion speed $v_c$ and direction $\theta$ on the field amplitude $h_x$ (up to $3.5 \times 10^{-3} J/g\mu_B$) is depicted in Fig. 5a, and the long-time skyrmion trajectories (up to 1.5 ns) at different field amplitudes are depicted in Fig. 5b. The frequency $\omega$ and phase $\varphi$ are fixed to be 105 rad/s and 0°, respectively. As expected, a larger field causes a more significant gyration of the skyrmion, and consequently leads to a more notable ratchet motion. The dependence of the ratchet motion speed on the field amplitude is in a power function trend as $v_c \propto h_x^\varepsilon$, with index $\varepsilon = 3.32$ is quite near the value of $m + n$. Note, a power function dependence of the ratchet motion speed on the field amplitude is a common feature of the ratchet motion of many soliton systems[37]. Meanwhile, the motion direction changes slightly with respect to the field amplitude. A decrease of the motion angle $\theta$ is observed at large fields and becomes more significant when the field is larger.

In additional to the field amplitude, the frequency also has a significant impact on the ratchet motion. In Fig. 5c, the ratchet motion speed $v_c$ and direction $\theta$ as functions of field frequency $f = \omega/2\pi$ are shown, with field amplitude $h_x$ and phase $\varphi$ being fixed to be $0.003 J/g\mu_B$ and 0°, respectively. The long-time skyrmion trajectories (up to 1.5 ns) at different frequencies are shown in Fig. 5d. It is found that the skyrmion ratchet motion speed has abnormal rumplings near the resonant frequency of the gyration mode

(Fig. 1d). This abnormal rumplings is believed to be caused by that fact that a series of modes (with different frequencies) rather than a single-frequency mode is excited in the skyrmion dynamics driven by the biharmonic field (we will discuss this in next section). Not only the amplitudes but also the phases of the excited modes are affected by the frequency of the biharmonic field. They together determine the ratchet motion speed. The effect of the frequency of the biharmonic field on the phases of the excited modes is complicated. This is also reflected in dependence of the ratchet motion direction on the field frequency. Contrast to the gentle effect of the field amplitude (Fig. 5a and b), the effect of field frequency on the ratchet motion direction is much more significant. In particular, a large change of motion direction occurs nearby the resonant frequency.

The ratchet motion speed $v_c$ and direction $\theta$ as functions of the phase $\varphi$ between the two harmonic field components are shown in Fig. 5e, and the long-time skyrmion trajectories (up to 1.5 ns) at different phases are shown in Fig. 5f. Here we fix the field amplitude to be $0.003 J/g\mu_B$, and the frequency to be 105 rad/s. Remarkably, with the phase changing from 0 to 360°, the skyrmion motion angle also has a full 360° rotation in CW direction. Note also that the ratchet motion speed shows a slight anisotropy along the angle direction, with two minimums at $\varphi = 90°$ and 270° (correspondingly, $\theta = 230°$ and 50°), and two maximums at $\varphi = 0°$ and 180° (correspondingly, $\theta = 320°$ and 140°). Therefore, one can readily use the field amplitude and frequency to tune the skyrmion ratchet motion speed, and use the phase to realize a 360° control of the skyrmion motion direction. This feature should be very useful in practice.

**D. Analysis of the skyrmion ratchet effect based on the Thiele's equation**

To understand the origin of the skyrmion ratchet effect, we further analyze in the frequency domain the dynamics of the skyrmion motion driven by biharmonic in-plane magnetic fields. The power spectra of the magnetic fields, the dissipative parameter $\mathcal{D}$, and the skyrmion coordinate component $x_c$ (result of $y_c$ is similar) for the two cases of magnetic fields, $h_x(t) = 0.003[\sin(\omega t) + \sin(3\omega t)]$ and $h_x(t) = 0.003[\sin(\omega t) + \sin(2\omega t)]$, are shown in Fig. 6a to c and 6d to f, respectively. The profiles of the dissipative parameter $\mathcal{D}$ in the time domain are also plotted in the inserts, and those of the magnetic fields and the skyrmion coordinate component $x_c$ are already shown in Fig. 3. It is clear that a series of modes with different frequencies rather than a single-frequency mode is excited in the skyrmion dynamics by the biharmonic field. Importantly, the excitation spectra of the dissipative parameter $\mathcal{D}$, as well as those of the coordinate $x_c$ show quite different features for the two cases. For the first case $(m, n) = (1, 3)$, only those modes with frequencies being even times of $\omega$, i.e., $2\omega, 4\omega, 6\omega, \ldots$, are excited in the spectrum of the dissipative parameter $\mathcal{D}$, and in contrast, the important modes of the skyrmion coordinate are those with frequencies that are odd times of $\omega$, i.e., $\omega, 3\omega, 5\omega, \ldots$. For the latter case with $(m, n) = (1, 2)$, frequencies that are even or odd times of $\omega$, i.e., $\omega, 2\omega, 3\omega, \ldots$, are all excited in both the spectrum of the dissipative parameter $\mathcal{D}$ and that of the skyrmion coordinate. Such a difference shows that the frequency overlapping of the excitation modes of the dissipative parameter and those of the skyrmion coordinate is the key to the skyrmion ratchet motion.

This is understandable, since such an overlapping would lead to a net dissipation force $\alpha \mathcal{D} \dot{\mathbf{R}}$ averaged over a period in Thiele's equation [Eq. (1)]. Here we rewrite

Thiele's equation in the component form as,

$$\begin{cases} -\mathcal{M}\dfrac{d^2x}{dt^2} - \mathcal{G}\dfrac{dy}{dt} - \alpha\mathcal{D}\dfrac{dx}{dt} - \mathcal{F}_x = 0 \\ -\mathcal{M}\dfrac{d^2y}{dt^2} + \mathcal{G}\dfrac{dx}{dt} - \alpha\mathcal{D}\dfrac{dy}{dt} - \mathcal{F}_y = 0 \end{cases} \qquad (8)$$

Here, $\mathbf{F}(t) = (\mathcal{F}_x(t), \mathcal{F}_y(t))$ is the biharmonic force related to the biharmonic magnetic field. In the following, we Suppose $\mathcal{F}_x(t) = f_1 \sin(m\omega t) + f_2 \sin(n\omega t + \varphi)$, $\mathcal{F}_y(t) = 0$, and that such a biharmonic force would cause a multi-mode excitation of the dissipative parameter $\mathcal{D}$. For simplicity, we assume the most important four modes of $\mathcal{D}$ are those at frequencies $2m\omega$, $2n\omega$, $(m-n)\omega$ and $(m+n)\omega$, so that we write

$$\begin{aligned}\mathcal{D}(t) = \mathcal{D}_0 &+ a_1 \sin(2m\omega t) + a_2 \sin(2n\omega t + 2\varphi) \\ &+ a_3 \sin\big((n-m)\omega t + \varphi'\big) + a_4 \sin\big((n+m)\omega t + \varphi'\big)\end{aligned} \qquad (9)$$

where $\varphi' = \varphi + \pi/2$, $\mathcal{D}_0 = 5.577\pi$ is the value of the dissipative parameter without excitation, and $a_1$, $a_2$, $a_3$ and $a_4$ are the amplitude of the four excited modes. The skyrmion dynamics under such a biharmonic force $\mathcal{F}_x(t)$ and excited modes of $\mathcal{D}$ can then be obtained by numerical solving of Thiele's equation. In the following, we set $\alpha = 0.1$, $\mathcal{M} = 0.13$ ns, $\mathcal{G} = -4\pi$, $\mathcal{D}_0 = 5.577\pi$, $a_1$=0.12, $a_2$=0.06, $a_1$=1.22, and $a_2$=0.026.

In Fig. 7, we depict two examples of long-time skyrmion trajectories (up to 2 ns) under biharmonic driving force $\mathcal{F}_x(t) = f_1 \sin(m\omega t) + f_2 \sin(n\omega t + \varphi)$ with $(m, n) = (1, 3)$ and $(m, n) = (1, 2)$ as predicted by Thiele's equation, with $f_1 = f_2 = 400$ a/s, $\omega/2\pi = 8$ GHz, and $\varphi$ varying from 0 to 360° by a step of 18°. One can see that, a ratchet motion is found for case $(m, n) = (1, 2)$ and the phase can realize a 360° control of the skyrmion motion direction, whereas a bound motion is found for case $(m, n) = (1, 2)$. We also find that the excited modes of the skyrmion coordinate do have a frequency overlapping with those of the dissipative parameter $\mathcal{D}$ for the case $(m, n) = (1, 2)$, whereas no overlapping occurs for case $(m, n) = (1, 3)$. This result is well consistent with our previous LLG simulation results. Therefore, a net dissipation force $\alpha \mathcal{D} \dot{\mathbf{R}}$ over a period due to an overlapping of the excitation modes of the dissipative parameter $\mathcal{D}$ and those of the skyrmion coordinate is indeed the key to the appearance of skyrmion ratchet motion.

**E. Discussion**

We would like to further point out that the solution of a ratchet motion driven by a biharmonic force based on Thiele's equation is not necessary to require a multi-mode spectrum of the dissipative parameter of $\mathcal{D}$. For example, for a biharmonic force with frequencies denoted by $(m, n)$, one can check that a solution of a ratchet motion is given by Thiele's equation even if $\mathcal{D}$ is assumed to have a single excitation mode at frequency that is the submultiple of the element frequency, e.g., $m\omega$ or $n\omega$. This is due to the fact that, for a driving force with a specific frequency, e.g., $m\omega$, if one assumes that the an excited mode of $\mathcal{D}$ is at frequency, e.g., $q\omega$, then a series of frequencies of the skyrmion coordinate $\mathbf{R}$ will be excited, including $m\omega$, $|m+q|\omega$, $|m-q|\omega$, $|m+2q|\omega$, $|m-2q|\omega$, …. The condition of ratchet motion is to have a frequency overlapping of the excitation modes of the dissipative parameter $\mathcal{D}$ and those of the skyrmion coordinate $\mathbf{R}$. So that there is a net dissipation force $\langle \alpha \mathcal{D} \mathbf{v} \rangle \neq 0$, and consequently $\langle \mathbf{v} \rangle \neq 0$ according to Thiele's equation. Thus once $\mathcal{D}$ has an excitation mode at frequency that is the submultiple of the element frequency, this condition is always satisfied.

In Table I, we list the dependence of the ratchet motion on the relation between the frequencies of the biharmonic field components quantified by (*m*, *n*), the excited modes of the dissipative parameter $\mathcal{D}$ predicted by LLG simulations, and the nontrivial single-modes or dual-modes (not exhaustive) of $\mathcal{D}$ that can lead to a ratchet motion predicted by Thiele's equation. From the Table, one can see that for fields with $m + n =$ odd, LLG simulations show that the excited spectrum of $\mathcal{D}$ covers all the frequencies which are integer times of the elemental frequency. Such an excitation of course covers those nontrivial single-modes or dual-modes of $\mathcal{D}$ that can lead to a ratchet motion as predicted by Thiele's equation, and thus a ratchet motion is observed. For cases $m + n =$ even, LLG simulations show that the excited spectrum of $\mathcal{D}$ covers only frequencies that are even times of the elemental frequency. Meanwhile the nontrivial single-modes of $\mathcal{D}$ that can lead to a ratchet motion as predicted by Thiele's equation are either *m* or *n*, which are both odd numbers. A bound motion is thus observed.

We emphasize that the ratchet effect revealed in this work should be distinguished in both the source application and the dynamics behind from those driven by oscillating driving forces as reported in previous works[33-37]. The ratchet motion, either driven by a biased oscillating magnetic field[33] or by oscillating drives in combination with substrate asymmetry[35,36], is intrinsically due to a spatial symmetry-breaking, as introduced by the field or by the substrate. For the ratchet motion driven by a tilted oscillating magnetic field[34], while the dissipation force should also play an important role in the net motion, such a net dissipation force relies on the co-excitation of the gyration mode caused by the in-plane field and the breathing mode caused by the out-of-plane field. The key to the ratchet motion under a high symmetric oscillating magnetic field gradient is also a time-dependent dissipation, but is caused by the coupling of the external field with the magnons[37]. In this work, the ratchet motion of skyrmion is driven by a biharmonic force with temporal symmetry-breaking. Such a temporal symmetry-breaking force gives rise to a multi-mode exactions of the dissipation parameter and the skyrmion coordinates, leading to a net dissipation force, and consequently, the ratchet motion.

At the end, we would like to present a more complicate skyrmion ratchet motion driven by pulsed in-plane magnetic fields as shown in Fig. 8. Here, the pulsed fields are repetitive sequences of alternating positive pulse (in magnitude of $0.003J/g\mu_B$ and over a time $TN_1$) and negative pulse (in magnitude $-0.003J/g\mu_B$ and over a time $TN_2$) with $TN_1 + TN_2 = TN = 60$ ps. It shows that ratchet motion occurs if $TN_1 \neq TN_2$ (i.e., the pulsed field has a biased component), and the ratchet motion is most significant when $TN_1$ is about half or double of $TN_2$. The skyrmion motion under such pulsed magnetic fields can be understood by regarding the pulsed magnetic fields as combinations of a series of harmonic magnetic fields and a biased magnetic field. Thus the driving force is sort of a mixing product of both spatial and temporal symmetry-breaking.

## IV. CONCLUSIONS

Micromagnetic simulation and analysis based on Thiele's equation are performed to study the skyrmion dynamics under biharmonic driving force. It shows that ratchet motion of skyrmion can be induced by a biharmonic in-plane magnetic field $h_x(t) = h_1\sin(m\omega t) + h_2\sin(n\omega t + \varphi)$, with a facile controllability of the ratchet motion speed and direction by tuning the field amplitude, frequency and phase, provided that *m* and *n* are two coprime integers such that *m* + *n* is odd, that is, when the field has a temporal symmetry-breaking. We propose that the ratchet motion is caused by an overlapping of the excitation spectra of the dissipation parameter and the skyrmion coordinate, which leads to the appearance of a net dissipation force averaged over time. The demonstration of the skyrmion ratchet effect enables further insight into the dynamic and soliton-like features of magnetic skyrmion, and its controllability should be useful in practice.

**ACKNOWLEDGMENTS**

This work was supported by the National Key Basic Research Program of China (No. 2015CB351905), NSFC (Nos. 11602310, 11474363, 11672339). Yue Zheng also thanks support from Special Program for Applied Research on Super Computation of the NSFC-Guangdong Joint Fund (the second phase), Fok Ying Tung Foundation, Guangdong Natural Science Funds for Distinguished Young Scholar and China Scholarship Council.

[1] U. K. Rößler, A. N. Bogdanov and C. Pfleiderer, Nature **442**, 797 (2006).
[2] S. Mühlbauer, B. Binz, F. Joinetz, C. Pfleiderer, A. Rosch, A. Neubauer, R. Georgii and P. Böni, Science **323**, 915 (2009).
[3] C. Pappas, E. Lelievre-Berna, P. Falus, P. M. Bentley, E. Moskvin, S. Grigoriev, P. Fouquet and B. Farago, Phys. Rev. Lett. **102**, 197202 (2009).
[4] C. Pfleiderer, T. Adams, A. Bauer, W. Biberacher, B. Binz, F. Birkelbach, P. Böni, C. Franz, R. Georgii, M. Janoschek, F. Jonietz, T. Keller, R. Ritz, S. Mühlbauer, W. Münzer, A. Neubauer, B. Pedersen and A. Rosch, J. Phys.: Condens. Matter **22**, 164207

(2010).

[5] N. Kanazawa, J. H. Kim, D. S. Inosov, J. S. White, N. Egetenmeyer, J. L. Gavilano, S. Ishiwata, Y. Onose, T. Arima, B. Keimer and Y. Tokura, Phys. Rev. B **86**, 134425 (2012).

[6] T. Tanigaki, K. Shibata, N. Kanazawa, X. Yu, Y. Onose, H. S. Park, D. Shindo and Y. Tokura, Nano Lett. **15**, 5438 (2015).

[7] X. Z. Yu, N. Kanazawa, Y. Onose, K. Kimoto, W. Z. Zhang, S. Ishiwata, Y. Matsui and Y. Tokura, Nat. Mater. **10**, 106 (2011).

[8] S. X. Huang and C. L. Chien, Phys. Rev. Lett. **108**, 267201 (2012).

[9] W. Münzer, A. Neubauer, T. Adams, S. Mühlbauer, C. Franz, F. Jonietz, R. Georgii P. Böni, B. Pedersen, M. Schmidt, A. Rosch and C. Pfleiderer, Phys. Rev. B **81**, 041203 (2010).

[10] X. Z. Yu, Y. Onose, N. Kanazawa, J. H. Park, J. H. Han, Y. Matsui, N. Nagaosa and Y. Tokura, Nature **465**, 901 (2010).

[11] K. Shibata, J. Iwasaki, N. Kanazawa, S. Aizawa, T. Tanigaki, M. Shirai, T. Nakajima, M. Kubota, M. Kawasaki, H. S. Park, D. Shindo, N. Nagaosa and Y. Tokura, Nat. Nanotechnol. **10**, 589 (2015).

[12] S. Seki, X. Z. Yu, S. Ishiwata and Y. Tokura, Science **336**, 198 (2012).

[13] S. Heinze, K. von Bergmann, M. Menzel, J. Brede, A. Kubetzka, R. Wiesendanger, G. Bihlmayer and S. Bülgel, Nat. Phys. **7**, 713 (2011).

[14] O. Boulle, J. Vogel, H. Yang, S. Pizzini, D. de Souza Chaves, A. Locatelli, T.O.Menteş, A. Sala, L. D. Buda-Prejbeanu, O. Klein, M. Belmegueani, Y. Roussigné, A. Stashkevich, S. M. Chérif, L. Aballe, M. Foerster, M. Chshiev, S. Auffret, I. M. Miron and G. Gaudin, Nat. Nanotechnol. **11**, 449 (2016).

[15] W. Jiang, P. Upadhyaya, W. Zhang, G. Yu, M. B. Jungfleisch, F. Y. Fradin, J. E. Pearson, Y. Tserkovnyak, K. L. Wang, O. Heinonen, S. G. E. te Velthuis and A. Hoffmann, Science **349**, 283 (2015).

[16] I. Dzyaloshinsky, J. Phys. Chem. Solids **4** 241 (1958).

[17] T. Moriya, Phys. Rev. **120** 91 (1960).

[18] W. Jiang, X. Zhang, G. Yu, W. Zhang, X. Wang, M. B. Jungfleisch, J. E. Pearson,

X. Cheng, O. Heinonen, K. L. Wang, Y. Zhou, A. Hoffmann and S. G. E. te Velthuis, Nat. Phys. **13**, 162 (2016).

[19] Kai Litzius, Ivan Lemesh, Benjamin Krüger, Pedram Bassirian, Lucas Caretta, Kornel Richter, F. Büttner, K. Sato, O. A. Tretiakov, J. Förster, R. M. Reeve, M. Weigand, I. Bykova, H. Stoll, G. Schütz, G. S. D. Beach and M. Kläui, Nat. Phys. **13**, 170 (2017).

[20] A. Neubauer, C. Pfleiderer, B. Binz, A. Rosch, R. Ritz, P. G. Niklowitz and P. Böni, Phys. Rev. Lett. **102**, 186602 (2009).

[21] N. Kanazawa, Y. Onose, T. Arima, D. Okuyama, K. Ohoyama, S. Wakimoto, K. Kakurai, S. Ishiwata and Y. Tokura, Phys. Rev. Lett. **106** 156603 (2011).

[22] M. Mochizuki, X. Z. Yu, S. Seki, N. Kanazawa, W. Koshibae, J. Zang, M. Mostovoy, Y. Tokura and N. Nagaosa, Nat. Mater. **13**, 241 (2014).

[23] C. Schütte and M. Garst, Phys. Rev. B **90**, 094423 (2014).

[24] N. Romming, C. Hanneken, M. Menzel, J. E. Bickel, B.Wolter, B. K. Von, A. Kubetzka and R.Wiesendanger, Science **341**, 636 (2013).

[25] N. Nagaosa and Y. Tokura, Nat. Nanotechnol. **8**, 899 (2013).

[26] X. Z. Yu, N. Kanazawa, W. Z. Zhang, T. Nagai, T. Hara, K. Kimoto, Y. Matsui, Y. Onose and Y. Tokura, Nat. Commun. **3**, 988 (2012).

[27] Y. Okamura, F. Kagawa, M. Mochizuki, M. Kubota, S. Seki, S. Ishiwata, M. Kawasaki, Y. Onose and Y. Tokura, Nat. Commun. **4**, 2391 (2013).

[28] J. Sampaio, V. Cros, S. Rohart, A. Thiaville and A. Fert, Nat. Nanotechnol. **8**, 839 (2013).

[29] R. Tomasello, E. Martinez, R. Zivieri, L. Torres, M. Carpentieri and G. Finocchio, Sci. Rep. **4**, 6784 (2014).

[30] Y. H. Liu, Y. Q. Li, and J. H. Han, Phys. Rev. B 87, 100402 (2013).

[31] K. Everschor, M. Garst, B. Binz, F. Jonietz, S. Mühlbauer, C. Pfleiderer and A. Rosch, Phys. Rev. B **86**, 054432 (2012).

[32] L, Kong and J. Zang, Phys. Rev. Lett. **111**, 067203 (2013).

[33] W. Wang, M. Beg, B. Zhang, W. Kuch and H. Fangohr, Phys. Rev. B **92** 020403 (2015).

[34] K. W. Moon, D. H. Kim, S. G. Je, B.S. Chun, W. Kim, Z.Q. Qiu, S. B. Choe and C. Hwang, Sci. Rep. **6**, 20360 (2016).

[35] C. Reichhardt, D. Ray and C. J. Olson Reichhardt, New J. Phys. **17**, 073034 (2015).

[36] X. Ma, C. J. Olson Reichhardt, and C. Reichhardt, Phys. Rev. B **95**, 104401 (2017)

[37] C. Psaroudaki and D. Loss, Phys. Rev. Lett. **120**, 237203 (2018).

[38] Y. Zolotaryuk and M. Salerno, Phys. Rev. E **73**, 066621 (2006).

[39] N. R. Quintero, J. A. Cuesta and R. Alvarez-Nodarse, Phys. Rev. E **81**, 030102 (2010).

[40] M. Mochizuki, Phys. Rev. Lett. 108, 017601 (2012).

[41] K. W. Moon, B. S. Chun, W. Kim, Z. Q. Qiu and C. Hwang, Phys. Rev. B **89**, 064413 (2014).

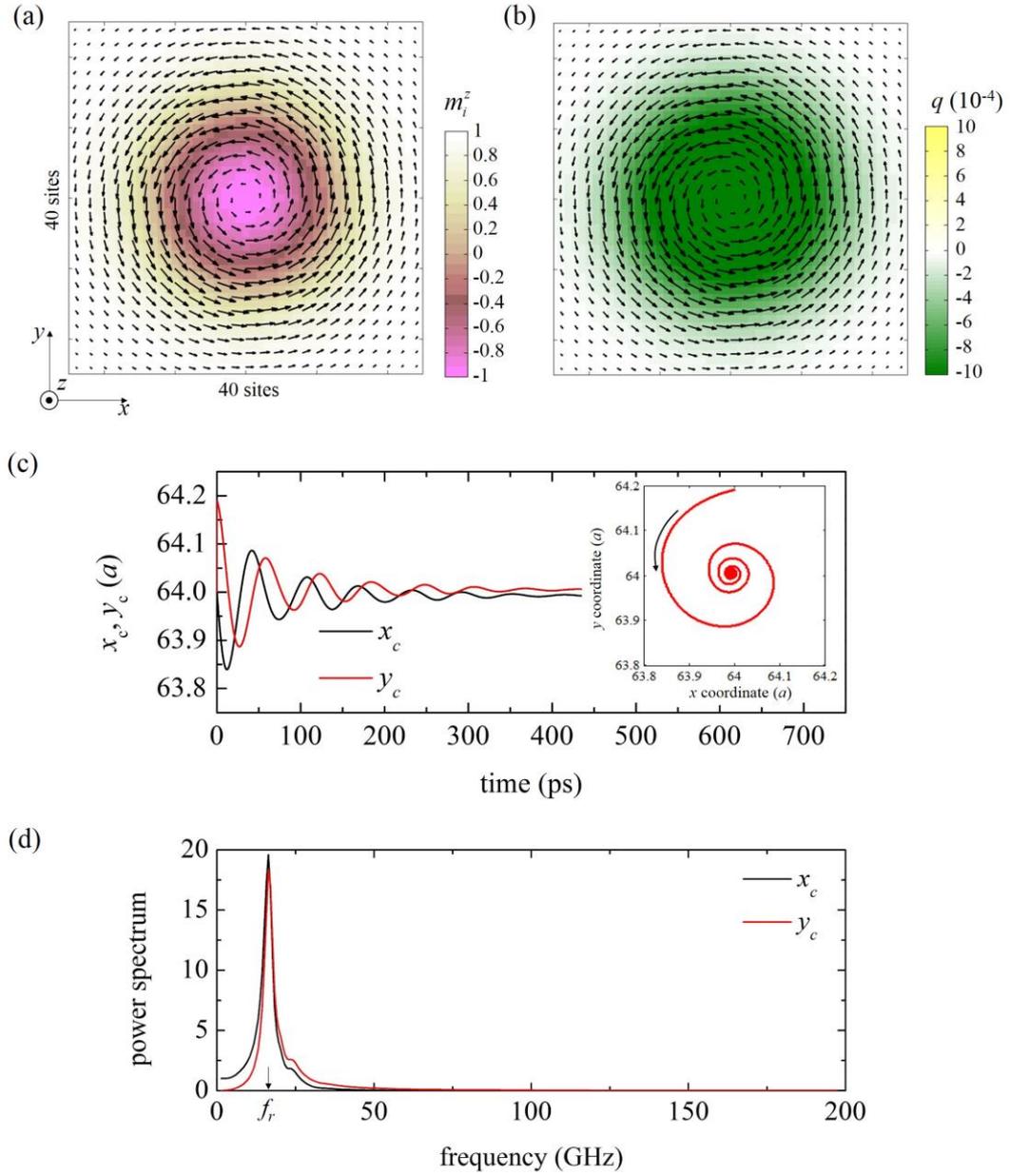

FIG. 1 Distributions of (a) the $z$-axis component magnetization $m_z$ and (b) topological charge density $q$ of the skyrmion at static state, with an arrow plot of the magnetization vectors projected onto the $xy$ plane at sites $(i_x, i_y)$ satisfying $\mathrm{mod}(i_x, 2) = \mathrm{mod}(i_x, 2) = 0$. (c) Time evolution of the skyrmion position $(x_c, y_c)$ after applying a $\delta$-function pulse of in-plane magnetic field $h_x(t) = h_0\delta(t)$ at $t = 0$. Trajectory of the skyrmion after excitation is depicted in the insert. (d) Power spectrum of the position of skyrmion $(x_c, y_c)$.

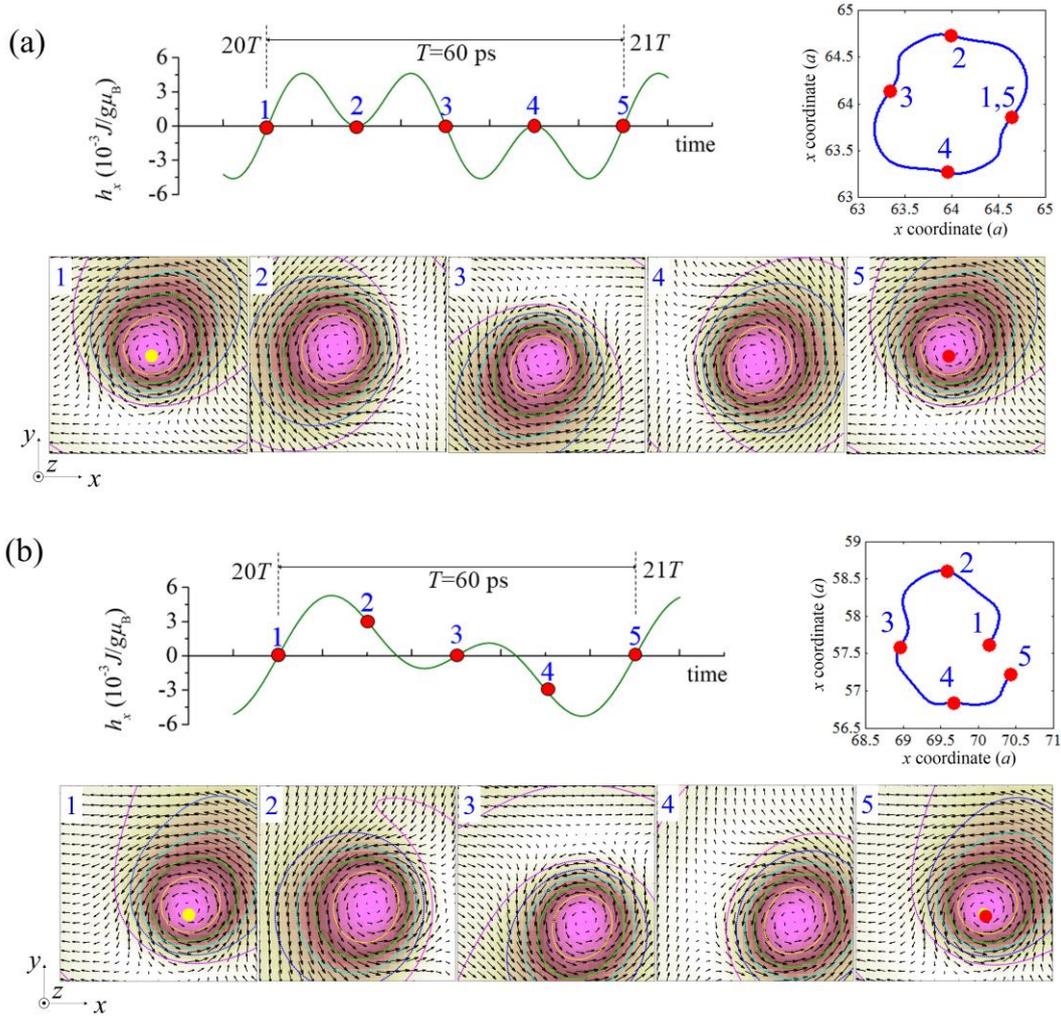

FIG. 2 Snapshots of skyrmion configuration under biharmonic in-plane magnetic fields (a) $h_x(t) = 0.003[\sin(\omega t) + \sin(3\omega t)]$ and (b) $h_x(t) = 0.003[\sin(\omega t) + \sin(2\omega t)]$. For each case, the snapshots are taken at five time points during a time period after the skyrmion reaches steady excitation, as labeled in the field profiles. The corresponding trajectory for each case is also shown in the right-top panel.

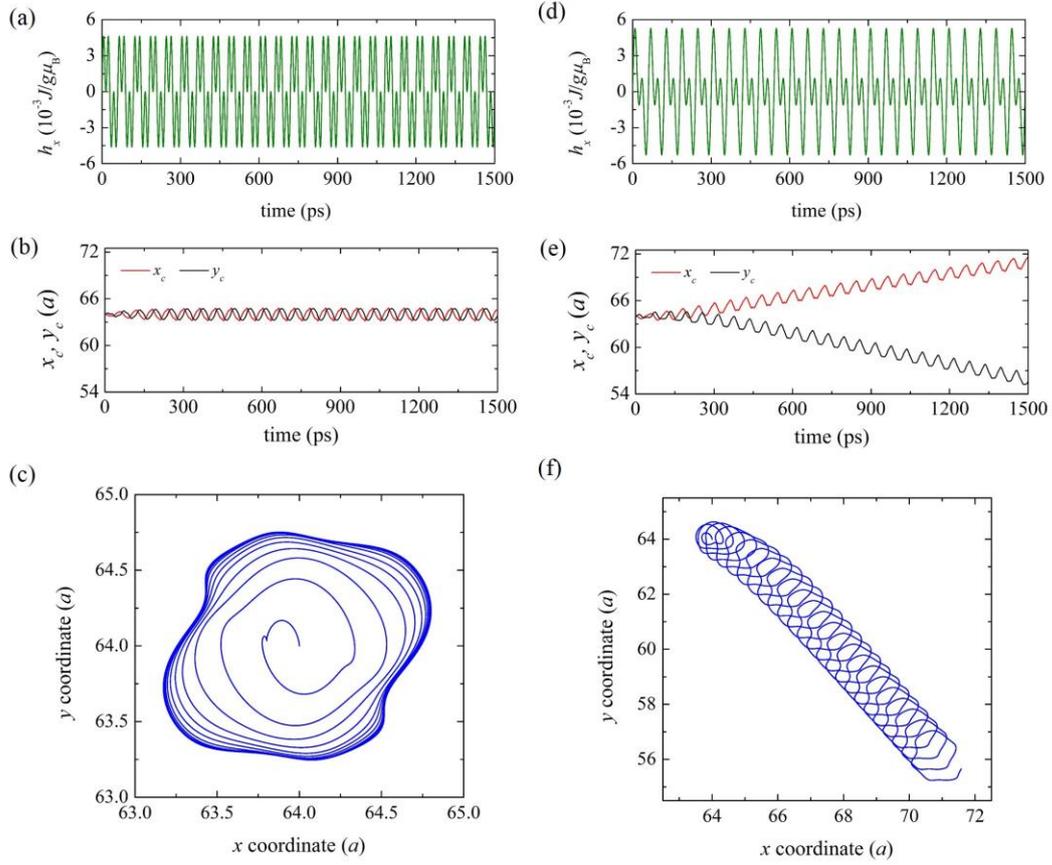

FIG. 3 Long-time skyrmion dynamics under biharmonic in-plane magnetic fields up to 1.5 ns. (a) to (c) Results for magnetic field $h_x(t) = 0.003[\sin(\omega t) + \sin(3\omega t)]$. (e) to (d) Results for magnetic field $h_x(t) = 0.003[\sin(\omega t) + \sin(2\omega t)]$. The field profiles, the time evolution curves of the skyrmion position coordinates $x_c$ and $y_c$, and the trajectories of the skyrmion for the two cases are shown in Fig. 3a to c, and Fig. 3d to f, respectively

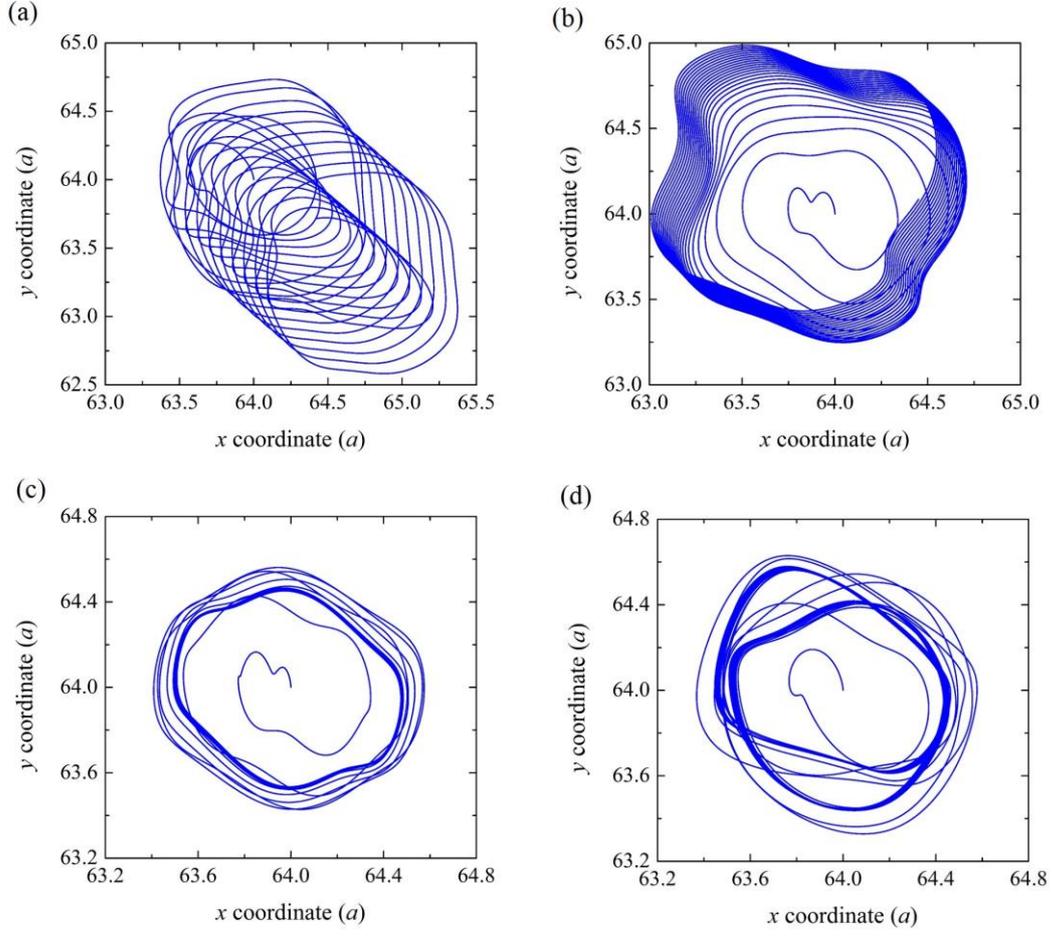

FIG. 4 Long-time skyrmion trajectories under different biharmonic in-plane magnetic fields in form of $h_x(t) = 0.003[\sin(m\omega t) + \sin(n\omega t)]$ up to 1.5 ns. (a) $(m, n) = (2, 3)$, (b) $(m, n) = (1, 4)$, (c) $(m, n) = (1, 5)$, and (d) $(m, n) = (2, 5)$.

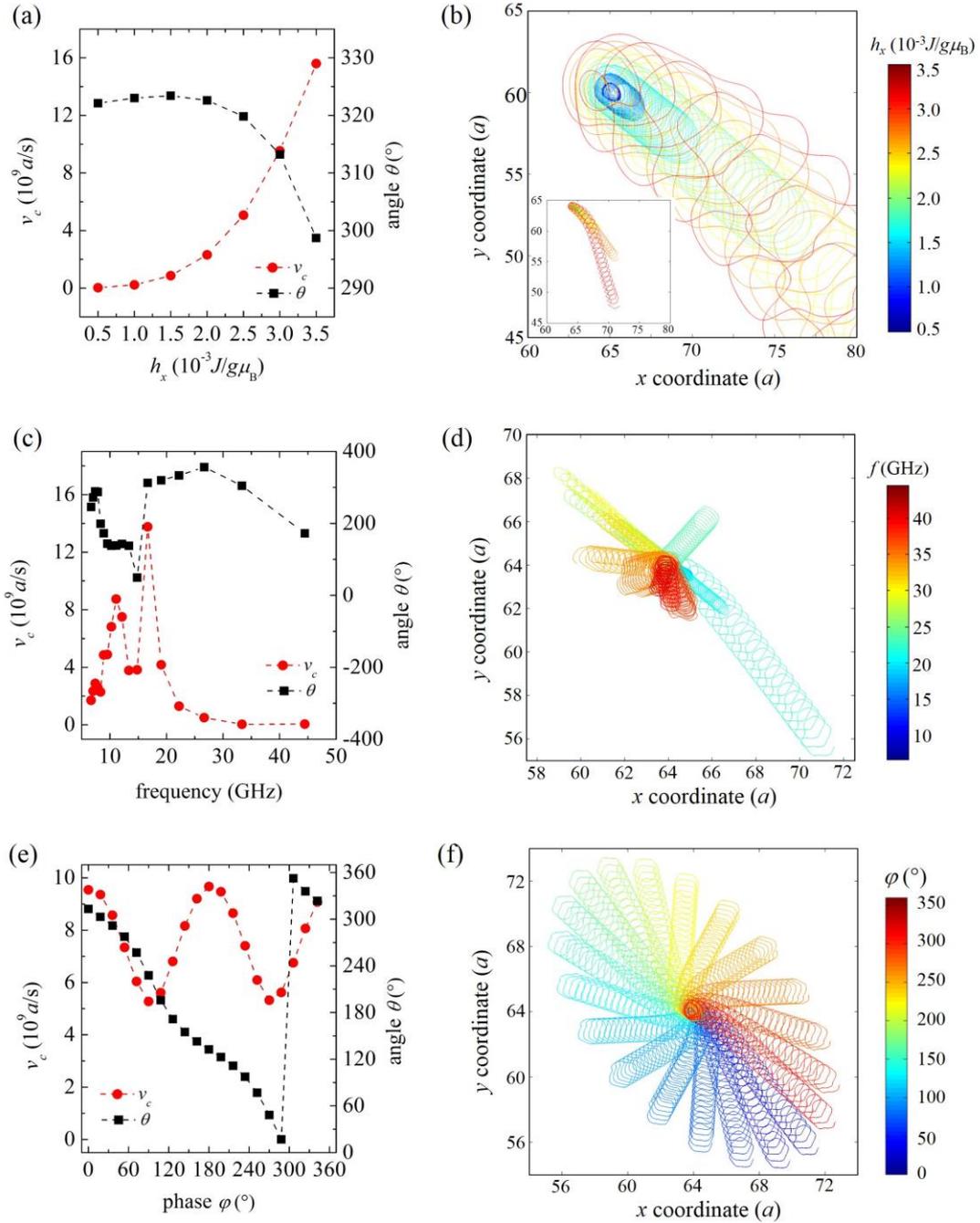

FIG. 5 Controllability of the skyrmion ratchet motion under biharmonic fields $h_x(t) = h_x[\sin(\omega t) + \sin(2\omega t + \varphi)]$. (a) Motion speed and direction as functions of field amplitude $h_x$. (b) Long-time skyrmion trajectories at different field amplitudes $h_x$. (c) Motion speed and direction as functions of field frequency $f=\omega/2\pi$. (d) Long-time skyrmion trajectories at different field frequencies $f$. (e) Motion speed and direction as functions of the phase $\varphi$. (f) Long-time skyrmion trajectories at different phases $\varphi$.

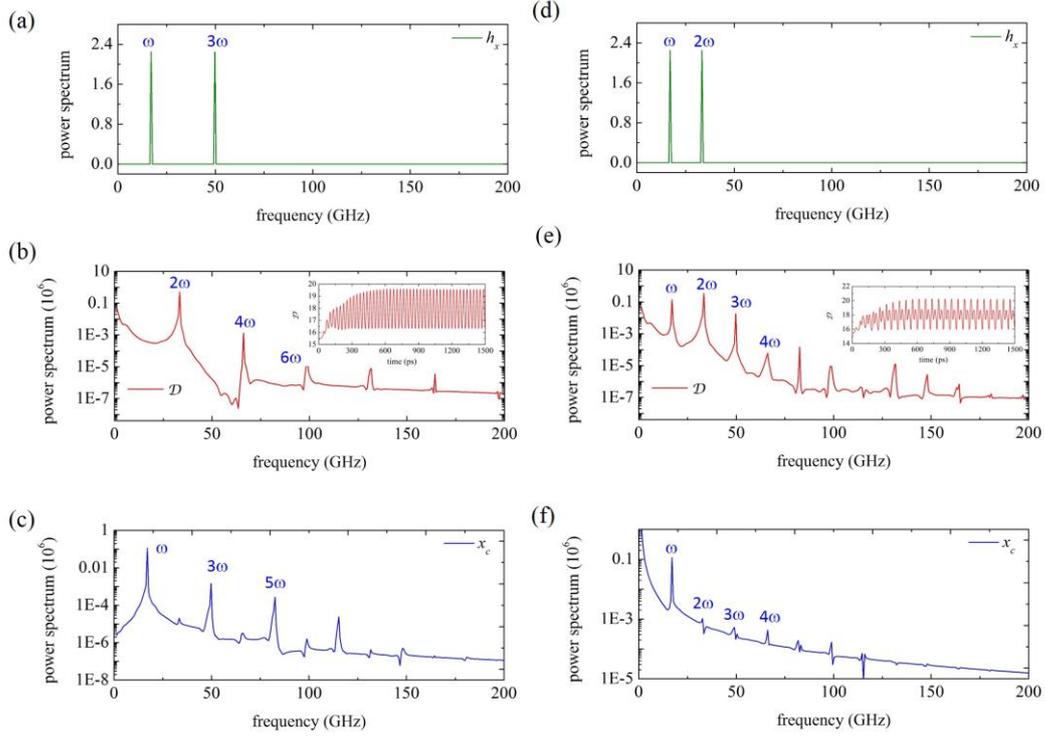

FIG. 6 Dynamics of the skyrmion motion driven by biharmonic in-plane magnetic fields in the frequency domain. Power spectra of the magnetic fields, the dissipative parameter $\mathcal{D}$, and the skyrmion coordinate component $x_c$ for the two cases of magnetic fields, (a) $h_x(t) = 0.003[\sin(\omega t) + \sin(3\omega t)]$ and $h_x(t) = 0.003[\sin(\omega t) + \sin(2\omega t)]$ are shown in Fig. 6a to c and 6d to f, respectively. The profiles of the dissipative parameter $\mathcal{D}$ in the time domain are also plotted in the inserts.

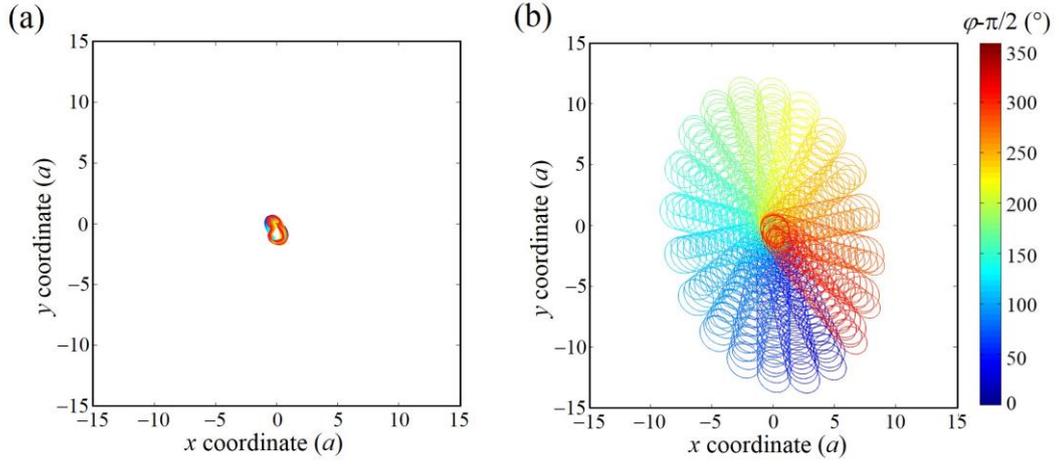

FIG. 7 Two examples of long-time skyrmion trajectories under biharmonic in-plane driving force $\mathcal{F}_x(t) = f_1 \sin(m\omega t) + f_2 \sin(n\omega t + \varphi)$ as predicted by Thiele's equation, with $f_1 = f_2 = 400$ $a$/s, $\omega/2\pi = 8$GHz, and $\varphi$ varying range from 0 to 360°. (a) $(m, n) = (1, 3)$ and (b) $(m, n) = (1, 2)$. The total time of each trajectory is 2 ns.

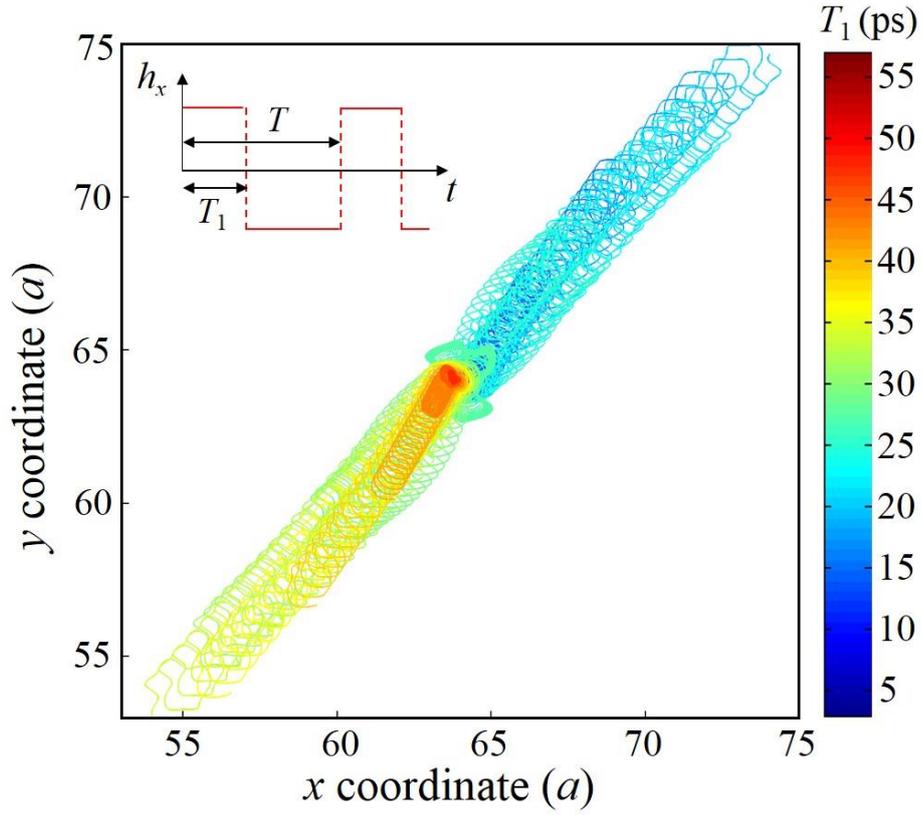

FIG. 8 Skyrmion ratchet motion driven by pulsed in-plane magnetic fields up to 1.5 ns. The pulsed fields are repetitive sequences of alternating positive pulse (in magnitude of $0.003J/g\mu_B$ and over a time $TN_1$) and negative pulse (in magnitude $-0.003J/g\mu_B$ and over a time $TN_2$) with $TN_1 + TN_2 = TN = 60$ ps.

TABLE I. Dependence of the ratchet motion on the relation between the frequencies of the biharmonic field components quantified by ($m$, $n$), the excited modes of the dissipative parameter $\mathcal{D}$ predicted by LLG simulations, and the nontrivial single-modes or dual-modes (not exhaustive) of $\mathcal{D}$ that can lead to a ratchet motion predicted by Thiele's equation.

| ($m$, $n$) | Excited modes of $\mathcal{D}$ | Nontrivial modes of $\mathcal{D}$ | Ratchet motion? |
|---|---|---|---|
| (1, 2) | 1, 2, 3, 4… (odd + even) | 1, 2, (2, 4)… | ✓ |
| (1, 3) | 2, 4, 6, 8… (even) | 1, 3… | ✗ |
| (1, 4) | 1, 2, 3, 4… (odd + even) | 1, 2, 4, (2, 8)… | ✓ |
| (1, 5) | 2, 4, 6, 8… (even) | 1, 5… | ✗ |
| (2, 3) | 1, 2, 3, 4… (odd + even) | 1, 2, 3, (4, 6)… | ✓ |
| (2, 5) | 1, 2, 3, 4… (odd + even) | 1, 2, 5, (4, 10)… | ✓ |